\documentclass[journal]{IEEEtran}
\usepackage{amsmath, amssymb, mathtools, relsize, paralist}

\newtheorem{theorem}{Theorem}
\newtheorem{corollary}[theorem]{Corollary}

\newcommand{ \defeq }{ \coloneqq }
\newcommand{ \mymod }{ \;\operatorname{mod}\, }

\begin{document}

\title{A Note on Parallel Asynchronous Channels\\with Arbitrary Skews}

\author{
        Mladen~Kova\v{c}evi\'c,~\IEEEmembership{Member,~IEEE}%
\thanks{Date: August 28, 2017.

        This work was supported by the Singapore Ministry of Education (MoE) Tier 2 grant
        ``Network Communication with Synchronization Errors: Fundamental Limits and Codes"
        (Grant number R-263-000-B61-112).

        The author is with the Department of Electrical \& Computer Engineering,
        National University of Singapore, Singapore 117583
				(email: mladen.kovacevic@nus.edu.sg).}%
       }%


\maketitle

\begin{abstract}
  A zero-error coding scheme of asymptotic rate $ \boldsymbol{\log_2 (1+\sqrt{5}) - 1} $
was recently described in \cite{engelberg+keren} for a communication channel composed
of parallel asynchronous lines satisfying the so-called no switch assumption.
We prove that this is in fact the highest rate attainable, i.e., the zero-error
capacity of this channel.
\end{abstract}%


\section{Introduction}


Let $ C^{(AAS)} $ denote the zero-error capacity of the Asynchronous channel
with an Arbitrary number of Skews (AAS); see \cite[Def.~1 and Def.~2]{engelberg+keren}.
Let $ C_{1,w} $ denote the zero-error capacity of the $ (1, w) $-channel; see
\cite[Sec.~IV]{engelberg+keren}.
It was recently shown in \cite[Cor.~11 and Cor.~15]{engelberg+keren} that
\begin{equation}
\label{eq:Caas}
  \log_2 \phi  \leq  C^{(AAS)}  \leq  \inf_{w \geq 2} C_{1,w} ,
\end{equation}
where $ \phi = \frac{1}{2}(1+\sqrt{5}) $ is the golden ratio.
We prove below that $ \inf_{w \geq 2} C_{1,w} = \log_2 \phi $, thus determining
the value $ C^{(AAS)} $ exactly.

The $ (1, w) $-channel, $ w \geq 2 $, is defined as follows.
The transmitter sends a signal of the form
\begin{equation}
\label{eq:input}
  S(t) = \sum_k  a_{k} \cdot p(t - kT) ,
\end{equation}
and the receiver obtains
\begin{equation}
\label{eq:output}
  R(t) = \sum_k  a_{k} \cdot p(t - kT - \tau - \tau_k) .
\end{equation}
Here $ a_k \in \{0, 1\} $ are binary symbols, $ T $ is the signaling interval, $ p(t) $
is a pulse of duration $ \leq T $, $ \tau $ is the propagation delay, and
$ \tau_k = \pm \frac{T}{2} $ is a random deviation of the delay of each pulse.%
\footnote{It will be evident from the proof of Theorem \ref{thm:C1w} that the zero-error
capacity of the $ (1, w) $-channel is unchanged if we assume that
$ \tau_k \in \big\{-\frac{T}{2}, 0, +\frac{T}{2} \big\} $.}
The channel is synchronous and the quantities $ T, p(t), \tau $ are known to the receiver.
Moreover, it is assumed that $ \tau_k = + \frac{T}{2} $ when $ k \equiv 1 \mymod w $ and
$ \tau_k = - \frac{T}{2} $ when $ k \equiv 0 \mymod w $.
In words, time is divided into groups consisting of $ w $ consecutive slots (i.e., signaling
intervals), and the pulses from one group cannot mix with pulses from a different group.
For example, in the group of time slots $ 1, \ldots, w $, the pulse in slot $ 1 $ (if
any) is ``shifted'' to the right by $ \frac{T}{2} $, the pulse in slot $ w $ (if any)
is shifted to the left by $ \frac{T}{2} $, while the pulses in slots $ 2, \ldots, w - 1 $
are shifted either to the left or to the right by $ \frac{T}{2} $, the choice being
random for each pulse; the same goes for the group $ w + 1, \ldots, 2 w $, and so on.

For the purpose of discussing the zero-error capacity of the channel just defined,
one can regard it \cite{engelberg+keren} as a memoryless channel with input alphabet
$ \{0, 1\}^w $.
The confusability graph \cite{shannon} corresponding to this channel is the graph
$ \Gamma_w $ with the vertex set $ \{0, 1\}^w $ and with an edge between any two strings
$ a, b \in \{0, 1\}^w $ that are confusable in the channel, meaning that they can produce
the same waveform \eqref{eq:output} at the channel output.
Evidently, $ a, b $ can only be confusable if they have the same Hamming weight.
Therefore, $ \Gamma_w $ has $ w + 1 $ connected components, one for each weight
$ h \in \{0, 1, \ldots, w\} $.

An \emph{adjacency reducing mapping} for a memoryless channel with input alphabet $ \mathcal A $
is a mapping $ f : {\mathcal A} \to {\mathcal A} $ with the property that if $ a, b $
are not confusable in the channel (i.e., not adjacent in the corresponding graph $ \Gamma $),
then $ f(a), f(b) $ are not confusable either.
The well-known result of Shannon \cite[Thm 3]{shannon} states that:
If there exists an adjacency reducing mapping such that its range, denoted $ f({\mathcal A}) $,
is an independent set in $ \Gamma $, then the zero-error capacity of the channel in
question equals $ \log_2 |f({\mathcal A})| $.
Clearly, in search for such mappings, one can consider each connected component of
$ \Gamma $ separately.

\section{The results}

\begin{theorem}
\label{thm:C1w}
  $ C_{1,w} = \frac{1}{w} \log_2 F_w $, where $ (F_w) $ is the Fibonacci sequence%
\footnote{The initial conditions of the Fibonacci sequence are usually taken to be
$ F_{0} = 0 $, $ F_1 = 1 $, so the sequence from Theorem \ref{thm:C1w} is in fact
the shifted Fibonacci sequence.
We ignore this fact here for notational simplicity.}
defined by $ F_{0} = 1 $, $ F_1 = 2 $, and $ F_w = F_{w-1} + F_{w-2} $ for $ w \geq 2 $.
\end{theorem}
\begin{IEEEproof}
  As mentioned above, $ \Gamma_w $ has $ w + 1 $ connected components, one for each weight
$ h $, so we can consider each of these components individually.
Denote by $ \{0, 1\}^w_h $ the set of all binary strings of length $ w $ and Hamming weight $ h $.
Each such string $ a = (a_1, \ldots, a_w) $ can be uniquely described by an $ h $-tuple
$ x^a = (x^a_1, \ldots, x^a_h) \in \mathbb{Z}^h $ satisfying $ 1 \leq x^a_1 < \cdots < x^a_h \leq w $,
where $ x^a_i $ represents the position of the $ i $'th $ 1 $ in the string $ a $.
For example, for $ a = (1, 0, 0, 1, 0) \in \{0,1\}^5_2 $, we have $ x^a = (1, 4) $.
For convenience, let us subtract from each such integer $ h $-tuple the vector $ (1, 2, \ldots, h) $
so that the resulting $ h $-tuples satisfy $ 0 \leq x^a_1 \leq \cdots \leq x^a_h \leq w - h $.
Denote
$ \Delta_{w-h}^h  \defeq  \big\{ x \in \mathbb{Z}^{h} : 0 \leq x_1 \leq \cdots \leq x_h \leq w-h \big\} $.
It is clear that $ \Delta_{w-h}^h $ is just a different representation of $ \{0, 1\}^w_h $
so we shall identify the two spaces with no risk of confusion.
Consequently, $ |\Delta_{w-h}^h| = \binom{w}{h} $.\\
Consider $ x, y \in \Delta_{w-h}^h $.
The key observation about the confusability graph $ \Gamma_w $ is the following:
$ x, y $ are confusable (i.e., $ (x, y) $ is an edge of $ \Gamma_w  $) if and only if
$ |x_i - y_i| \leq 1 $ for \emph{every} $ i \in \{1, \ldots, h\} $, i.e., if and only if
$ \left\| x - y \right\|_\infty \leq 1 $.
This is because each pulse of the transmitted waveform is shifted by either
$ +\frac{T}{2} $ or $ -\frac{T}{2} $ by definition, so
\begin{inparaenum}
\item[1.)]
if the $ i $'th pulse of one input waveform is two or more slots away from the
$ i $'th pulse of another input waveform, then these two waveforms cannot produce
the same output, and
\item[2.)]
if the $ i $'th pulse of one input waveform is at most one slot away from the
$ i $'th pulse of another input waveform, for every $ i \in \{1, \ldots, h\} $, then
these two waveforms can always produce the same output.
\end{inparaenum}\\
For a non-negative integer $ x_i $ define
\begin{equation}
\label{eq:fxi}
  f(x_i)  \defeq  2 \left\lfloor \frac{x_i}{2} \right\rfloor  =
  \begin{cases}
	  x_i      , & \text{if}\  x_i\  \text{is even} \\
    x_i - 1  , & \text{if}\  x_i\  \text{is odd} .
  \end{cases}
\end{equation}
Abusing the notation slightly define also the mapping
$ f : \Delta_{w-h}^h \to \Delta_{w-h}^h $ by $ f(x)  \defeq  \big(f(x_1), \ldots, f(x_h)\big) $.
We claim that $ f $ satisfies the conditions of \cite[Thm 3]{shannon}.
Indeed, it follows from \eqref{eq:fxi} that if $ |x_i - y_i| \geq 2 $, then
$ |f(x_i) - f(y_i)| \geq 2 $ as well.
In other words, if $ x, y $ are not confusable, then $ f(x), f(y) $ are not confusable
either, which means that $ f $ is an adjacency reducing mapping.
Furthermore, the range of $ f $ is an independent set in $ \Gamma_w $, i.e., every two
elements of $ f(\Delta_{w-h}^h) $ are non-confusable.
This is because $ f(\Delta_{w-h}^h) $ contains only $ h $-tuples with even entries, and
so for any $ x, y \in f(\Delta_{w-h}^h) $, $ x \neq y $, there must exist a coordinate
$ i $ where $ |x_i - y_i| \geq 2 $.
This proves that $ f $ satisfies the conditions of \cite[Thm 3]{shannon} and, therefore,
the zero-error capacity of the $ (1, w) $-channel equals
$ \frac{1}{w} \log_2 \sum_{h=0}^w |f(\Delta_{w-h}^h)| $.\\
To complete the proof we need to show that $ \sum_{h=0}^w |f(\Delta_{w-h}^h)| = F_w $.
To that end we first note that $ f(\Delta_{w-h}^h) $ can be written in the form
$ 2 \cdot \Delta_{d}^h $, where $ d = \lfloor \frac{w-h}{2} \rfloor $.
Since $ \Delta_{d}^h $ represents the set of binary strings of length $ d + h $ and
weight $ h $, we get
$ |f(\Delta_{w-h}^h)| = |\Delta_{d}^h| = \binom{h + \lfloor \frac{w-h}{2} \rfloor}{h} $.
One can now check directly that the sequence
$ S_w \defeq \sum_{h=0}^w \binom{h + \lfloor \frac{w-h}{2} \rfloor}{h} $
\linebreak\newpage\noindent
satisfies the recurrence $ S_w = S_{w-1} + S_{w-2} $.
As for the initial conditions, there is only one binary string of length $ w = 0 $
(empty string), and there are two binary strings of length $ w = 1 $ and they are
non-confusable.
Therefore, $ S_w = F_w $.
\end{IEEEproof}

The mapping $ f $ from the above proof, and its generalizations, were used in \cite{kp, kst}
to construct optimal zero-error codes for shift and timing channels.

\begin{corollary}
\label{thm:monotonic}
  $ \inf_{w \geq 2} C_{1,w} = \lim_{w \to \infty} C_{1,w} = \log_2 \phi $, where
$ \phi = \frac{1}{2}(1+\sqrt{5}) $ is the golden ratio.
\end{corollary}
\begin{IEEEproof}
  Theorem \ref{thm:C1w} implies $ \lim_{w \to \infty} C_{1,w} = \log_2 \phi $,
so it is enough to show that $ C_{1,w} \geq \log_2 \phi $, or equivalently that
$ F_w \geq \phi^w $, for every $ w \geq 2 $.
Solving the recurrence for $ F_w $ we find that
$ F_w = \alpha \phi^{w} + (1 - \alpha) (-\phi^{-1})^{w} $, where
$ \alpha = \frac{ 4 + 2\sqrt{5} }{ 5 + \sqrt{5} } \approx 1.17 $.
The statement $ F_w \geq \phi^w $ is therefore equivalent to\linebreak
$ (\alpha - 1) ( \phi^{w} - (-\phi^{-1})^{w} ) \geq 0 $,
which is easily verified to be true since $ \phi^{-1} < 1 $.
\end{IEEEproof}

\vspace{5.5mm}
\section*{Acknowledgment}

The author would like to thank Vincent Y. F. Tan for reading a preliminary
version of this work and providing several helpful comments.

\vspace{5mm}

%

\end{document}